\documentclass[aps,twocolumn,superscriptaddress,tightenlines]{revtex4}
\usepackage{epsfig,graphicx,times}
\usepackage{amstext}
\usepackage{amsmath}
\usepackage{amssymb}
\usepackage{graphicx}
\usepackage{latexsym}
\usepackage{bm}
\usepackage[colorlinks,citecolor=blue, linkcolor=blue,hyperindex,CJKbookmarks,dvipdfm]{hyperref}

\begin{document}

\title{Phononic Josephson oscillation and self-trapping with two-phonon exchange interaction}
\author{Xun-Wei Xu}
\email{davidxu0816@163.com}
\affiliation{Department of Applied Physics, East China Jiaotong University, Nanchang,
330013, China}
\author{Ai-Xi Chen}
\email{aixichen@ecjtu.edu.cn}
\affiliation{Department of Physics, Zhejiang Sci-Tech University, Hangzhou, 310018, China}
\affiliation{Department of Applied Physics, East China Jiaotong University, Nanchang,
330013, China}
\author{Yu-xi Liu}
\affiliation{Institute of Microelectronics, Tsinghua University, Beijing 100084, China}
\affiliation{Tsinghua National Laboratory for Information Science and Technology
(TNList), Beijing 100084, China}
\date{\today }

\begin{abstract}
We propose a bosonic Josephson junction (BJJ) in two nonlinear mechanical
resonator coupled through two-phonon exchange interaction induced by quadratic
optomechanical couplings. The nonlinear dynamic equations and effective
Hamiltonian are derived to describe behaviors of the BJJ. We show that
the BJJ can work in two different dynamical regimes: Josephson oscillation
and macroscopic self-trapping. The system can transfer from one regime to
the other one when the self-interaction and asymmetric parameters exceed their
critical values. We predict that a transition from Josephson oscillation to
macroscopic self-trapping can be induced by the phonon damping in the
asymmetric BJJs. Our results opens up a way to demonstrate BJJ with
two-phonon exchange interaction and can be applied to other systems,
such as the optical and microwave systems.
\end{abstract}

\maketitle

%\email{aixichen@ecjtu.edu.cn}

%\email{yuxiliu@tsinghua.edu.cn}

\section{Introduction}

Bosonic Josephson junction (BJJ), a bosonic analog of the superconducting
Josephson junction, was first proposed and observed in two weakly coupled
Bose-Einstein condensates~\cite%
{MilburnPRA97,SmerziPRL97,RaghavanPRA99,MarinoPRA99,GiovanazziPRL00,SarchiPRA08,BackhausNat98,AlbiezPRL05,LevyNat07,TrenkwalderNPY16}
to study macroscopic tunneling. After that BJJ has also been studied
both theoretically and experimentally in other nonlinear bosonic
systems, such as coupled nonlinear optical cavities~\cite%
{ACJiPRL09,DidierPRB11,LagoudakisPRL10,AbbarchiNPy13,VoronovaPRL15,RahmaniSR16,LarsonPRA11,JHTengJPB12}
and nanomechanical resonators~\cite{BarzanjehPRA16}. One important
application of BJJ is to serve as a quantum interference device~\cite%
{GeraceNPy09}. As a two-mode Bose-Hubbard model, BJJ also offers a simple
platform to explore quantum many-body dynamics~\cite{HartmannLPR08}.

In contrast to all hitherto realized BJJs, where two nonlinear bosonic
systems are coupled by hopping of single bosons, we here propose a BJJ in
two nonlinear mechanical modes coupled through two-phonon exchange interaction.
The Bose-Hubbard model with atom-pair tunneling~\cite{FollingNat07,ZollnerPRL08,JQLiangPRA09,RubeniPRA17,PietraszewiczPRA12} or two-photon exchange~\cite%
{AlexanianPRA10,AlexanianPRA11,YLDongPRA12,HardalJOSAB12,HardalJOSAB14,TaianarX17,HWangarX17} has been studied for years.
However, the realization of two-phonon exchange interaction in the mechanical systems is still lack of effective method.

Recently, multi-mode optomechanical system~\cite{AspelmeyerARX13}, that multiple mechanical resonators are coupled to a single cavity mode via radiation pressure or optical gradient forces, provides us an appropriate platform to realize nonlinear phononic interaction mediated by the cavity mode~\cite{LudwigPRA10,SeokPRA12,SeokPRA13,BuchmannPRA15,XWXuPRA13}.
We find that two-phonon exchange interaction can be induced by coupling two
mechanical modes to a common cavity mode through quadratic optomechanical
interactions. An effective Hamiltonian for two nonlinear mechanical modes
with two-phonon exchange interaction is obtained by adiabatically
eliminating the cavity mode. We show that the transition between Josephson
oscillation and macroscopic self-trapping (MST) can be observed by tuning
the parameters blow (or above) certain critical values.

Different from the BJJs with single-boson hopping interaction, where four
distinct modes are predicted~\cite{SmerziPRL97,RaghavanPRA99,MarinoPRA99},
i.e., zero-phase mode, running-phase mode, $\pi$-phase oscillations, and $%
\pi $-phase self-trapping, whereas in our system, there are three distinct
modes, i.e., zero-phase mode, $\pi/2$-phase oscillations, and running-phase
mode. In addition, a dynamic transition from Josephson oscillation to MST
induced by phonon damping is predicted for asymmetric BJJ with two-phonon
exchange interaction, which is very different from the previous
theoretical predictions~\cite{MarinoPRA99,RahmaniSR16} and experimental
observations~\cite{AbbarchiNPy13}.

The paper is organized as follows. In Sec.~II, we derive an effective Hamiltonian for two
nonlinear mechnical modes coupled through two-phonon exchange interaction from an multi-mode quadratic optomechanical system with two nonlinear mechanical modes and one cavity mode.
In Sec.~III, an effective Hamiltonian for BJJ are obtained from the two
nonlinear mechnical modes coupled through two-phonon exchange interaction.
The behavior of the BJJ in the non-self-interacting and linear regimes is discussed in Sec.~IV.
In Sec.~V, we study the dynamic behaviors for the symmetric BJJ.
The effective potential for BJJ is shown in Sec.~VI.
We study the effect of the phonon damping on the dynamic behavior of
the asymmetric BJJ in Sec.~VII.
Finally, we summarize results in Sec.~VIII.

\section{two-phonon exchange interaction}

\begin{figure}[tbp]
\includegraphics[bb=50 250 566 428, width=8.5 cm, clip]{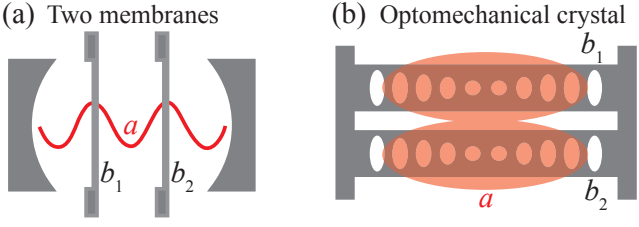}
\caption{(Color online) Schematic diagram of two nonlinear mechanical modes (%
$b_{1}$ and $b_{2}$) coupled to a common cavity mode ($a$) through quadratic
optomechanical interactions: (a) two partly reflective nonlinear membranes
in a Fabry-Perot cavity; (b) optomechanical crystal.}
\label{fig1}
\end{figure}

As schematically shown in Fig.~\ref{fig1}, we study a system that two nonlinear mechanical modes are coupled to a
common cavity mode with quadratic optomechanical couplings. Such system can be realized by either
two partly reflective nonlinear membranes in a Fabry-Perot cavity~\cite%
{ThompsonNat08}, optomechanical crystal~\cite{EichenfieldNat09,MeenehanPRX15,ParaisoPRX15}, or other systems.
The Hamiltonian of these systems can be written as
\begin{eqnarray}
H &=&\sum_{i=1,2}\left[ \omega _{i}^{(0)}b_{i}^{\dag
}b_{i}+U_{i}^{(0)}b_{i}^{\dag }b_{i}^{\dag }b_{i}b_{i}+g_{i}a^{\dag }a\left(
b_{i}^{\dag }+b_{i}\right) ^{2}\right]   \notag \\
&&+\omega _{c}a^{\dag }a+\left( \Omega a^{\dag }e^{-i\omega _{d}t}+\mathrm{%
H.c.}\right) ,
\end{eqnarray}%
where $a$ and $a^{\dag }$ are the annihilation and creation operators of the
cavity mode with frequency $\omega _{c}$, $b_{i}$ and $b_{i}^{\dag }$ ($i=1,2
$) are the annihilation and creation operators of the $i$th nonlinear
mechanical mode with frequency $\omega _{i}^{(0)}$ and nonlinearity strength
$U_{i}^{(0)}$, and $g_{i}$ is the quadratic optomechanical coupling strength
between the cavity mode and the $i$th mechanical mode. The cavity mode is
driven by an external field with the strength $\Omega $ and frequency $\omega _{d}
$.

The dynamics of the mechanical oscillators and cavity mode can be described
by quantum Langevin equations. After considering the dissipations and within the mean-field approximation, we can obtain the dynamical equations
\begin{equation}
\frac{d}{dt}b_{i}=-\left( \frac{\gamma _{i}}{2}+i\omega _{i}^{(0)}\right)
b_{i}-i2U_{i}^{(0)}b_{i}^{\dag }b_{i}b_{i}-i2g_{i}a^{\dag }a\left(
b_{i}^{\dag }+b_{i}\right) ,
\end{equation}%
\begin{equation}
\frac{d}{dt}a=-\left\{ \frac{\kappa }{2}+i\left[ \Delta
_{c}+\sum_{i=1,2}g_{i}\left( b_{i}^{\dag }+b_{i}\right) ^{2}\right] \right\}
a-i\Omega ,
\end{equation}%
for $i=1,2$, with the damping rates of the cavity mode $\kappa $ and
mechanical modes $\gamma _{i}$ $(i=1,2)$.
Here, $\Delta _{c}\equiv \omega _{c}-\omega
_{d}$ is the detuning between the cavity mode and driving field.
To solve the above nonlinear
dynamical equations, we can write each operator as the sum of its steady-state
value and the time-dependent term: $a\rightarrow \alpha +a$ and $%
b_{i}\rightarrow \beta _{i}+b_{i}$, where $\alpha $ and $\beta _{i}$ are the
steady-state values of the system. When $g_{i}\geq 0$, the
steady-state values are
\begin{equation}
\alpha =\frac{-i2\Omega }{\kappa +i2\Delta _{c}},
\end{equation}%
\begin{equation}
\beta _{i}=0.
\end{equation}%
Under the assumption that the external driving is strong (i.e., $|\alpha |\gg 1$), $%
\left\vert \alpha \right\vert ^{2}\gg \left\langle a^{\dag }a \right\rangle$, and frequency shift induced by the quadratic
optomechanical couplings is small, $\Delta _{c}\gg \left\langle \sum_{i=1,2}g_{i}\left(
b_{i}^{\dag }+b_{i}\right) ^{2} \right\rangle$, the dynamical equations for the
time-dependent terms are given by
\begin{eqnarray}
\frac{d}{dt}b_{i} &=&-\left( \frac{\gamma _{i}}{2}+i\omega _{i}\right)
b_{i}-i2U_{i}^{(0)}b_{i}^{\dag }b_{i}b_{i} -i2g_{i}\left\vert \alpha \right\vert ^{2}b_{i}^{\dag }  \notag \\
&&-i2g_{i}\left( \alpha a^{\dag }+\alpha ^{\ast }a+ a^{\dag }a \right) \left( b_{i}^{\dag
}+b_{i}\right),
\end{eqnarray}%
\begin{equation}
\frac{d}{dt}a=-\left( \frac{\kappa }{2}+i\Delta _{c}\right)
a-i\sum_{i=1,2}g_{i}\alpha \left( b_{i}^{\dag }+b_{i}\right) ^{2},
\end{equation}%
where $\omega _{i}=\omega _{i}^{(0)}+2g_{i}\left\vert \alpha \right\vert ^{2}
$.

After introducing the slowly varying amplitudes $\widetilde{a}\equiv
ae^{\left( i\Delta _{c}+\kappa /2\right) t}$ and $\widetilde{b}_{i}\equiv
b_{i}e^{i\omega _{i}t}$, under the rotating-wave approximation (by keeping the terms with low frequencies $ \Delta
_{i}\equiv \Delta _{c}-2\omega _{i} $ and
neglecting oscillating terms with high frequencies, e.g. $\omega _{i}$, $\Delta _{c}$, etc) with $\left\vert \Delta
_{i}\right\vert \ll \left\{
\omega _{i},\Delta _{c}\right\} $, we have%
\begin{equation}\label{addEq8}
\frac{d\widetilde{b}_{i}}{dt}=-i2U_{i}^{(0)}\widetilde{b}_{i}^{\dag }%
\widetilde{b}_{i}\widetilde{b}_{i}-i2G_{i}^{\ast }\widetilde{a}\widetilde{b}%
_{i}^{\dag }e^{-\left( i\Delta _{i}+\kappa /2\right) t}-\frac{\gamma _{i}}{2}%
\widetilde{b}_{i},
\end{equation}%
\begin{equation}
\frac{d\widetilde{a}}{dt}=-i\sum_{i=1,2}G_{i}\widetilde{b}_{i}^{2}e^{\left(
i\Delta _{i}+\kappa /2\right) t},
\end{equation}%
where $G_{i}=g_{i}\alpha $ is the effective optomechanical coupling
strength.
The expression of $\widetilde{a}$ can be obtained as
\begin{equation}
\widetilde{a}=-i\sum_{i=1,2}\int_{-\infty }^{t}G_{i}\widetilde{b}_{i}^{2}e^{\left(
i\Delta _{i}+\kappa /2\right) \tau}d\tau.
\end{equation}%
Under the assumption that the damping rate of the cavity mode is
much larger than the effective optomechanical couplings $G_{j}$ and the damping
rates of the mechanical modes, i.e., $\kappa \gg \left\{ G_{j},\gamma
_{i}\right\} $,
the evolution of $\widetilde{b_{i}}$ is much slower than $\widetilde{a}$, so $\widetilde{b_{i}}$ can be taken out of the integrals,
then we have~\cite{JahnePRA09,XWXuPRA15}
\begin{equation}\label{addEq10}
\widetilde{a} =-i\sum_{i=1,2}\frac{2G_{i}}{\kappa +i2\Delta _{i}}\widetilde{b}_{i}^{2}e^{\left(
i\Delta _{i}+\kappa /2\right) t}.
\end{equation}%
After adiabatically eliminating the cavity mode by substituting Eq.~(\ref{addEq10}) and $b_{i}\equiv
\widetilde{b}_{i}e^{-i\omega _{i}t}$ into Eq.~(\ref{addEq8}), the dynamical equations for the mechanical modes $b_{i}$ become
\begin{equation}
\frac{db_{1}}{dt}=-\left( \frac{\gamma _{1}}{2}+i\omega _{1}\right)
b_{1}-i2U_{1}b_{1}^{\dag }b_{1}b_{1}+i2J_{1}b_{2}b_{2}b_{1}^{\dag },
\end{equation}%
\begin{equation}
\frac{db_{2}}{dt}=-\left( \frac{\gamma _{2}}{2}+i\omega _{2}\right)
b_{2}-i2U_{2}b_{2}^{\dag }b_{2}b_{2}+i2J_{2}b_{1}b_{1}b_{2}^{\dag },
\end{equation}%
where the effective nonlinearity strength $U_{i}=U_{i}^{(0)}-\left\vert
G_{i}\right\vert ^{2}/\left( \Delta _{i}-i\kappa /2\right) $ and effective
two-phonon exchange coupling strengths $J_{1}=-g_{1}g_{2}\left\vert \alpha
\right\vert ^{2}/\left( \Delta _{2}-i\kappa /2\right) $ and $J_{2}=-g_{1}g_{2}\left\vert \alpha
\right\vert ^{2}/\left( \Delta _{1}-i\kappa /2\right) $ can be controlled by tuning the strength $\Omega $ and frequency $%
\omega _{d}$ of the external field. We choose $%
\{\left\vert \Delta _{1}-\Delta _{2}\right\vert ,\kappa \}\ll
\left\vert \Delta _{i}\right\vert $, so that $U_{i}\approx U_{i}^{(0)}-g_{i}^{2}\left\vert \alpha \right\vert
^{2}/\Delta _{i}$ and $J\approx -g_{1}g_{2}\left\vert \alpha \right\vert
^{2}/\Delta _{1}\approx -g_{1}g_{2}\left\vert \alpha \right\vert ^{2}/\Delta
_{2}$, and a Hermitian Hamiltonian can be obtained as (without considering the damping terms)
\begin{equation}\label{addEq14}
H_{\mathrm{eff}}=\sum_{i=1,2}\left( \omega _{i}b_{i}^{\dag
}b_{i}+U_{i}b_{i}^{\dag }b_{i}^{\dag }b_{i}b_{i}\right) +J\left( b_{1}^{\dag
}b_{1}^{\dag }b_{2}b_{2}+b_{2}^{\dag }b_{2}^{\dag }b_{1}b_{1}\right),
\end{equation}
which describes a model for two
nonlinear mechnical modes coupled through two-phonon exchange interaction.

\section{Bosonic Josephson junction}

From Eq.~(\ref{addEq14}), we can verify that, when the effect of mechanical damping can be
neglected, the total phonon population $N_{T}=n_{1}+n_{2}$ is constant,
where $n_{i}$ is the phonon population in the $i$th mechanical mode. For
large phonon numbers, i.e., $N_{T}\gg 1$, the operators of the mechanical
modes can be treated as classical quantities,
\begin{equation}
b_{i}=\sqrt{n_{i}}e^{i\theta _{i}},
\end{equation}%
where $\theta _{i}$ is the phase. By introducing the population imbalance $%
z\equiv\left( n_{1}-n_{2}\right) /N_{T}$ and the phase difference $\phi \equiv \theta
_{2}-\theta _{1}$ between the two mechanical modes, the dynamics of the mechanical modes can be rewritten as the
nonlinear equations
\begin{equation}
\frac{dz}{dt}=\left( 1-z^{2}\right) \sin 2\phi ,  \label{eq6}
\end{equation}%
\begin{equation}
\frac{d\phi }{dt}=\Delta +gz-z\cos 2\phi ,  \label{eq7}
\end{equation}%
where the time has been rescaled as $2JN_{T}t\rightarrow t$, and the
dimensionless parameters are $g=(U_{1}+U_{2})/2J$, $\Delta =\Delta
_{0}+\Delta _{u}$ with $\Delta _{0}=(\omega _{1}-\omega _{2})/2JN_{T}$ and $%
\Delta _{u}=(U_{1}-U_{2})/2J$. We can see that these nonlinear dynamical
equations are invariant under the transformation $\Delta \rightarrow -\Delta
$, $\phi \rightarrow -\phi +\pi /2$\ and $g\rightarrow -g$.
%We notice that both the
%asymmetric parameter $\overline{\Delta }$ and the self-interaction parameter $g$ are
%independent of $N_{T}$.

We can consider $z$ and $\phi $ as two canonically conjugate variables, then an
effective Hamiltonian (derived from the above equation with $dz/dt=-\partial
H_{J}/\partial \phi $ and\ $d\phi /dt=\partial H_{J}/\partial z$) for BJJ is
obtained as%
\begin{equation}  \label{eq8}
H_{J}=\Delta z+\frac{g}{2}z^{2}+\frac{1}{2}\left( 1-z^{2}\right) \cos 2\phi .
\end{equation}%
The BJJ tunneling current is defined by%
\begin{equation}
I\equiv \frac{N_{T}}{2}\frac{dz}{dt}=JN_{T}^{2}\left( 1-z^{2}\right) \sin
2\phi .
\end{equation}

\section{Non-self-interacting and linear regimes}

Before the detailed analysis of the BJJ with numerical solutions, here we
consider the behavior of the system in the non-self-interacting and linear
regimes. For symmetric BJJ without self-interaction, i.e., $\Delta =g=0$,
the dynamical equation for $z$ is obtained as%
\begin{equation}
\frac{d^{2}z}{dt^{2}}=-2z\left( 1-z^{2}\right) .
\end{equation}%
This can yield a hamonic oscillation for $z$ only in the limit $\left\vert
z\right\vert \ll 1$ with frequency%
\begin{equation}
\omega _{0}=2\sqrt{2}JN_{T}.
\end{equation}
In this case the BJJ tunneling current $I$ is an alternating current (AC) with frequency $2\sqrt{2}JN_{T}$.

For symmetric BJJ ($\Delta =0$) with $g<1$, in the linear limit ($\left\vert
z\right\vert \ll 1$ and $\left\vert \phi \right\vert \ll 1$), the dynamical
equation for $z$ is given by%
\begin{equation}
\frac{d^{2}z}{dt^{2}}=2\left( g-1\right) z.
\end{equation}%
$z$ oscillates hamonic with frequency%
\begin{equation}
\omega _{L}=2\sqrt{2\left( 1-g\right) }JN_{T}.
\end{equation}
Then the frequency of the AC current $I$ become $2\sqrt{2\left( 1-g\right) }JN_{T}$.

Still in the linear limit ($\left\vert z\right\vert \ll 1$ and $\left\vert
\phi \right\vert \ll 1$) with $g<1$, if the BJJ is asymmetric with parameter
$\Delta \gg \left( g-\cos 2\phi \right) z$, then we have%
\begin{equation}
\phi =\phi \left( 0\right) +\Delta t,
\end{equation}%
\begin{equation}
z=z\left( 0\right) -\frac{1}{2\Delta }\cos \left[ 2\phi \left( 0\right)
+2\Delta t\right] .
\end{equation}%
$z$ oscillates harmonically with frequency%
\begin{equation}
\omega _{\mathrm{ac}}=2\Delta .
\end{equation}%
The BJJ tunneling current is given by%
\begin{equation}
I=JN_{T}^{2}\sin \left[ 2\phi \left( 0\right) +2\Delta t\right] .
\end{equation}%
An AC current $I$ is produced in the asymmetric BJJ working in the linear
limit.

\section{Symmetric BJJ}

\begin{figure}[tbp]
\includegraphics[bb=47 389 539 583, width=8.5 cm, clip]{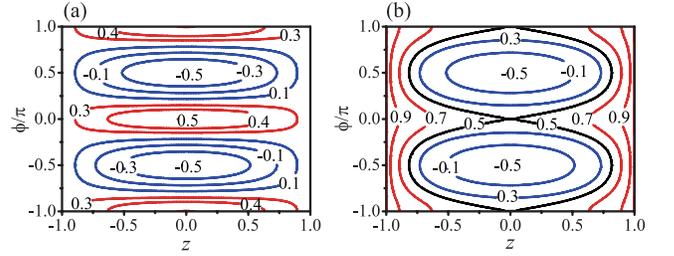}
\caption{(Color online) Energy contours of a symmetric bosonic Josephson
junction versus $z$ and $\protect\phi$, for (a) $g=0.5$ and (b) $g=2.0$.}
\label{fig2}
\end{figure}

For a symmetric BJJ, i.e., $\Delta =0$, the Hamiltonian in Eq.~(\ref{eq8})
becomes
\begin{equation}  \label{eq18}
H_{J}=\frac{g}{2}z^{2}+\frac{1}{2}\left( 1-z^{2}\right) \cos 2\phi .
\end{equation}%
Figure~\ref{fig2} shows the energy contours of a symmetric BJJ for different
values of self-interaction parameter $g$. We can find that the location of
the energy minima, maxima, and saddle points crucially depends upon the
self-interaction parameter $g$. For $g<1$ (strong two-phonon exchange
coupling, i.e, $U_{i}<J$), the minima are at $\left[ z,\phi \right] =\left[
0,\left( m+1/2\right) \pi \right] $ ($m$ is an integer) and the maxima
settle in $\left[ z,\phi \right] =\left[ 0,m\pi \right] $, whereas for $g>1$
(strong nonlinearities, i.e, $U_{i}>J$), the minima are still at $\left[
z,\phi \right] =\left[ 0,\left( m+1/2\right) \pi \right] $, while $\left[
z,\phi \right] =\left[ 0,m\pi \right] $ become saddle points. This
transition of the point $\left[ z,\phi \right] =\left[ 0,m\pi \right] $ from
a local maximum to a saddle point is a manifestation of the transition of
the Josephson oscillation regime to the self-trapping regime.

For a given value of the initial population imbalance $z\left( 0\right) $,
if the self-interaction parameter $g$ exceeds a critical value $g_{c}$, the
populations become macroscopically self-trapped with $\left\langle
z\right\rangle \neq 0$. This corresponds to the macroscopic self-trapping (MST) condition for
\begin{equation}  \label{eq19}
H_{0}\equiv H_{J}\left( z\left( 0\right),\phi \left( 0\right) \right) >\frac{%
1}{2},
\end{equation}
and the critical self-interaction parameter for MST is%
\begin{equation}  \label{eq20}
g_{c}=\left\{1-\left[ 1-z\left( 0\right) ^{2}\right] \cos \left[ 2\phi \left(
0\right) \right] \right\}\frac{1}{z\left( 0\right) ^{2}}.
\end{equation}

\begin{figure}[tbp]
\includegraphics[bb=20 251 575 626, width=8.5 cm, clip]{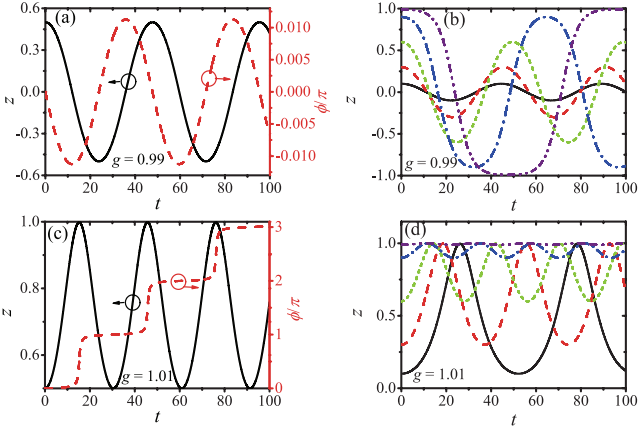}
\caption{(Color online) Population imbalance $z$ and phase difference $%
\protect\phi/\protect\pi$ as functions of the rescaled time $t$ for $g=0.99$
in (a) and (b), $g=1.01$ in (c) and (d). The initial imbalance in (a) and
(c) is $z(0)=0.5$. In (b) and (d), the imbalance takes the initial values $%
0.1$ (black solid), $0.3$ (red dash), $0.6$ (green dot), $0.9$ (blue dash-dot), $0.99$ (purple dash-dot-dot). The other parameters are $\Delta =0$ and
$\protect\phi (0)=0$.}
\label{fig3}
\end{figure}

Figure~\ref{fig3} describes the time evolution of population imbalance $z$
for different values of self-interaction parameter
$g$. Figures~\ref{fig3}(a) and \ref{fig3}(c) show the transition from the
Josephson oscillation to the MST regime at $g=1$, for the specific 
initial conditions $[z(0),\phi(0)] = [0.5,0]$. In Fig.~\ref{fig3}(a), where $g<1$, $z$ and $\phi$ oscillate around $[z,\phi] = [0,0]$, which corresponding to the zero-phase mode. In Fig.~\ref{fig3}(c), where $g>1$, $\langle z\rangle =0$ and $\phi$ increases monotonously, which corresponding to the running-phase mode. The transition behavior for $%
\phi(0) = 0$ at $g=1$ is independent of the initial value of
the population imbalance, as shown in Figs.~\ref{fig3}(b) and \ref{fig3}(d).
We can also see this clearly from Eq.~(\ref{eq20}): $g_{c}=1$ for $\phi(0) =
0$, which is independent of the population imbalance.

On the other hand, from Eq.~(\ref{eq19}), when $g>1$ remains constant and initial value $\phi
\left( 0\right) \neq m\pi$ ($m$ is an integer), there is a critical population imbalance $z_{c}$ for the initial value of the population imbalance $z\left( 0\right)$ as
\begin{equation}
z_{c}=\sqrt{\frac{1-\cos \left[ 2\phi \left( 0\right) \right] }{g-\cos \left[
2\phi \left( 0\right) \right] }}.
\end{equation}
\begin{figure}[tbp]
\includegraphics[bb=8 336 590 715, width=8.5 cm, clip]{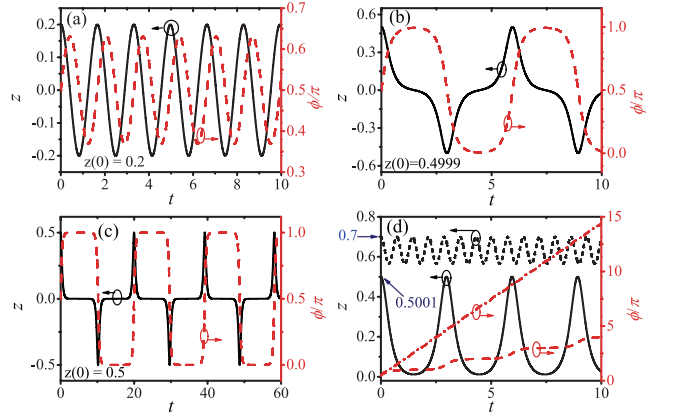}
\caption{(Color online) Population imbalance $z$ and phase difference $%
\protect\phi/\protect\pi$ as functions of the rescaled time $t$ for $\protect%
\phi (0)=\protect\pi/2$, $g=7$ and $\Delta =0$. The initial imbalance $z(0)$
takes the initial values (a) $0.2$, (b) $0.4999$, (c) $0.5$, (d) $0.5001$
(black solid and red dash curves) and $0.7$ (black dot and red dash-dot curves).}
\label{fig4}
\end{figure}

Figure~\ref{fig4} shows the transition from the Josephson oscillation to the
MST regime at $z(0)=z_{c}=0.5$, for the specific initial
conditions $[\phi(0),g] = [\pi/2,7]$. For $z(0)<0.5$, an increase of $z(0)$
adds higher harmonics to the sinusoidal oscillations, and the period of such
oscillations increases with $z(0)$, as shown in Figs.\ref{fig4}
(a)-(c). Meanwhile, $\phi$ oscillates around the point $\phi= \pi/2$, which corresponding to the $\pi/2$-phase oscillations. MST occurs when $z\left(0\right) >z_{c}$ as shown in Fig.~\ref{fig4}(d).
Moreover, for $z(0)>0.5$, the period and the amplitude of the MST
oscillations decrease with $z(0)$, i.e., $z$ becomes more localized with
high oscillation frequency for larger values of $z(0)$.

\section{Potential for BJJ}

In this section, we employ the alternative approach of examining the effective potential for the BJJ.
One can use the energy $H_{J}$ of Eq.~(\ref{eq18}) to describe the system in terms of an equation of motion for
a classical particle moving in a potential $W(z)\equiv H_{J}-\left( dz/dt\right) ^{2}$~\cite{RaghavanPRA99} with coordinate $z$ and total energy $H_{J}$. For symmetric BJJ ($\Delta=0$), the potential $W(z)$ is
obtained as
\begin{eqnarray}
W\left( z\right) &=&H_{0}+4H_{0}^{2}-1  \notag \\
&&+2\left( 1-2H_{0}g\right) z^{2}+\left( g^{2}-1\right) z^{4}
\end{eqnarray}%
with the conserved energy $H_{0}=H_{J}$. It is clear that if $H_{0}>1/2$, we
will have $W\left( z=0\right) >H_{0}=H_{J}$ and MST sets in.

\begin{figure}[tbp]
\includegraphics[bb=34 372 552 569, width=8.5 cm, clip]{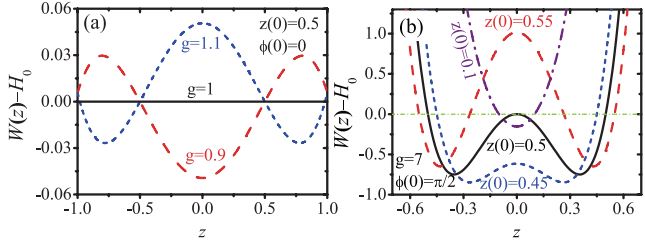}
\caption{(Color online) The potential $W(z)-H_{0}$ is plotted as a function of $z$ for
initial phase difference $\protect\phi(0)=0$ in (a) and $\protect\phi(0)=%
\protect\pi/2$ in (b) with $\Delta =0$. In (a) $g$ takes the values $0.9$
(red dash), $1.0$ (black solid) and $1.1$ (blue dot) with the initial
population imbalance $z(0)=0.5$. In (b) initial population imbalance $z(0)$
takes the values $0.1$ (purple dash-dot), $0.55$ (red dash), $0.5$ (black
solid) and $0.45$ (blue dot) with $g=7$.}
\label{fig5}
\end{figure}

Figure~\ref{fig5} displays the potential $W(z)$ for (a) $\phi(0)=0$%
, (b) $\phi(0)=\pi/2$. For a given value of initial conditions $%
[z(0),\phi(0)]=[0.5,0]$, in Fig.~\ref{fig5}(a), the the increase of the the value $g$,
$W(z)$ is changed from a single (red dash curve) to a double (blue dot curve)
well and the changeover occurs at the critical point $g=1$ (black solid
horizontal line). It is worth noting that the potential is flat at the
critical point $g=1$ corresponding to steady state for the population
imbalance $z$. For a given value of $[\phi(0),g]=[\pi/2,7]$, in Fig.~\ref%
{fig5}(b), with the increase of $z(0)$, $W(z)$ is changed from a parabolic
(purple dash-dot curve) to a double well and the changeover occurs at the
point $z(0)=\sqrt{1/g}$. As the parameter $g$ increases, the oscillations
become anharmonic and the system is in the Josephson regime [also see Fig.~%
\ref{fig4}(b) and \ref{fig4}(c)]. For $z(0)>0.5$ the total energy is smaller
than the potential barrier (red dash curve), forcing the particle to become
localized in one of the two wells.

\begin{figure}[tbp]
\includegraphics[bb=67 283 520 584, width=8.5 cm, clip]{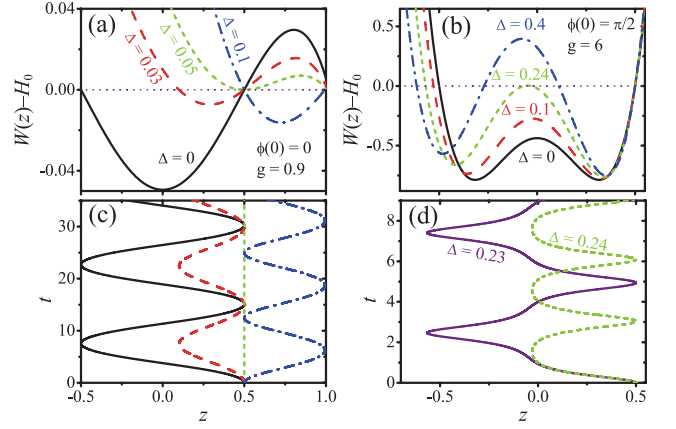}
\caption{(Color online) $z$ potential $W(z)-H_{0}$ plotted against $z$ for [$%
\protect\phi(0)=0$, $g=0.9$] in (a) and [$\protect\phi(0)=\protect\pi/2$, $%
g=6$] in (b) with $z(0)=0.5$ and $\Delta $ taking different values. $z$ is
plotted as a function of the rescaled time $t$ for [$\protect\phi(0)=0$, $%
g=0.9$] in (c) and [$\protect\phi(0)=\protect\pi/2$, $g=6$] in (d) with $%
z(0)=0.5$ and $\Delta $ taking different values.}
\label{fig6}
\end{figure}

If the BJJ is asymmetric with $\Delta \neq 0$, then the potential $W(z)$ is
given by
\begin{eqnarray}
W\left( z\right) &=&H_{0}+4H_{0}^{2}-1-8H_{0}\Delta z  \notag \\
&&+2\left( 1+2\Delta ^{2}-2H_{0}g\right) z^{2}  \notag \\
&&+4\Delta gz^{3}+\left( g^{2}-1\right) z^{4}.
\end{eqnarray}%
The potential is asymmetric because there are two terms with odd powers of $z$ in $W\left(
z\right) $. We plot $W(z)-H_{0}$ for different
$\Delta $ in Figs.~\ref{fig6}(a) and (b). The corresponding dynamical
evolution of $z$ is shown in Figs.~\ref{fig6}(c) and (d). For $%
\Delta =0$ the potential $W(z)-H_{0}$ is symmetric and $z$ oscillates around
$\langle z\rangle=0$ for $g<g_{c}$. When $\Delta $ is increased, asymmetric energy is added
to the potential, corresponding to $\langle z\rangle \neq
0 $. If the asymmetric parameter $\Delta $ exceeds a critical value $\Delta
_{c}$ [green dot curves, $\Delta _{c}=0.05$ in Figs.~\ref{fig6}(a) and (c), $%
\Delta _{c}\approx 0.24$ in Figs.~\ref{fig6}(b) and (d)], the system moves
into the MST regime.

\section{Damping induced transition}

We now consider the effect of the phonon damping on the dynamic behavior of
the asymmetric BJJ. For simplicity, we assume that the two nonlinear
mechanical modes have the same damping rates, i.e., $\gamma _{0}\equiv
\gamma _{1}=\gamma _{2} $. The dynamical equations in the presence of phonon
damping are given by
\begin{equation}  \label{eq24}
\frac{dz}{dt}=\left( e^{-\gamma t}-z^{2}\right) \sin 2\phi -\frac{1}{2}%
\gamma z,
\end{equation}%
\begin{equation}  \label{eq25}
\frac{d\phi }{dt}=\Delta _{0}+\Delta _{u}e^{-\frac{\gamma }{2}t}+gz-z\cos
2\phi ,
\end{equation}%
where $\gamma =\gamma _{0}/(JN_{T})$ is the dimensionless damping parameter,
and $N_{T}$ is the total phonon population at the initial time. The damping
parameter $\gamma $ is inversely proportional to the initial total phonon
population $N_{T}$. In order to suppress the effect of the phonon damping,
one effective way is to enhance the total phonon population $N_{T}$ in the
initial time.

\begin{figure}[tbp]
\includegraphics[bb=19 367 572 565, width=8.5 cm, clip]{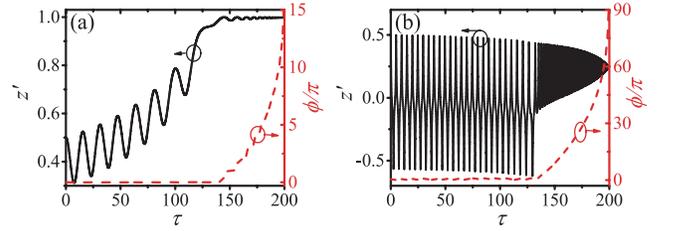}
\caption{(Color online) Population imbalance $z^{\prime }$ and phase
difference $\protect\phi$ as functions of the rescaled time $\protect\tau$
for $\protect\gamma=0.01$, in (a) with $z^{\prime }(0)=0.5$, $\protect\phi %
(0)=0$, $g=0.9$, $\Delta_0=0.03$, $\Delta_u=0.01$, and in (b) with $%
z^{\prime }(0)=0.5$, $\protect\phi (0)=\protect\pi/2$, $g=6$, $\Delta_0=0.22$%
, $\Delta_u=0.01$.}
\label{fig7}
\end{figure}

Equations~(\ref{eq24}) and (\ref{eq25}) can be rewritten as the dynamical
equations of the effective population imbalance $z^{\prime }\equiv ze^{\frac{%
\gamma }{2}t}$ and $\phi $ as
\begin{equation}  \label{eq26}
\frac{dz^{\prime }}{d\tau }=\left( 1-z^{\prime 2}\right) \sin 2\phi ,
\end{equation}%
\begin{equation}  \label{eq27}
\frac{d\phi }{d\tau }=\frac{2\Delta _{0}}{2-\gamma \tau }+\Delta
_{u}+gz^{\prime }-z^{\prime }\cos 2\phi ,
\end{equation}%
where the rescaled time $\tau $ is defined by $\tau \equiv \frac{2}{\gamma }\left(
1-e^{-\frac{\gamma }{2}t}\right) $ for $t\in \left[ 0,+\infty \right) $ and $%
\tau \in \left[ 0,2/\gamma\right) $. Equations~(\ref{eq26}) and (\ref{eq27}) are
the same as Eqs.~(\ref{eq6}) and (\ref{eq7}) but with $\Delta _{0}$ replaced
by $2\Delta _{0}/(2-\gamma \tau) $. This means that the asymmetric of the
system are enhanced as time goes on. When the asymmetric parameter $%
\Delta^{\prime }=2\Delta _{0}/(2-\gamma \tau )+\Delta _{u}$ exceeds the
critical value $\Delta _{c}$, the system has a transition from the Josephson
oscillation into the MST. Figure~\ref{fig7} shows the time evolution of $%
z^{\prime }$ and $\phi $ as functions of the rescaled time $\tau $ in the
presence of phonon damping, i.e., $\gamma =0.01$. It shows that the system
works in the Josephson oscillation regime at the beginning with $\Delta
_{0}+\Delta _{u}<\Delta _{c}$, and then moves into the MST regime when
$\tau >[2-2\Delta _{0}/(\Delta _{c}-\Delta_{u})]/\gamma $.

\section{Conclusions}

In summary, we have proposed a BJJ in two nonlinear mechanical resonator
coupled through two-phonon exchange interaction. Two dynamic regimes of
Josephson oscillation and MST are predicted, and the system can transfer
from one regime to the other one when the self-interaction and asymmetric
parameters exceed their critical values. A transition, from Josephson
oscillation to MST induced by the phonon damping, can
be observed in the asymmetric BJJs.
The measurement of the dynamic behaviors of the mechanical resonators could be realized by transferring
the mechanical signals into electric signals through piezoelectric effect~\cite{OkamotoNP13,PHuangPRL13}, or into optical signals through auxiliary optomechanical couplings~\cite{BarzanjehPRA16,XWXuPRA15,VitaliPRL07,PalomakiNature13,CohenNature15}.
Our results open a way to investigate
interferometer and Bose-Hubbard model with two-phonon exchange interactions
in optomechanical systems. Similarly, BJJ based on two-boson exchange
interaction can also be realized in the optical and microwave systems~\cite%
{AlexanianPRA10,AlexanianPRA11,YLDongPRA12,HardalJOSAB12,HardalJOSAB14,TaianarX17,HWangarX17}.

%\section{Acknowledgement}
\vskip 2pc \leftline{\bf Acknowledgement}

X.-W.X. is supported by the National Natural Science Foundation of China
(NSFC) under Grants No.11604096 and the Startup Foundation for Doctors of
East China Jiaotong University under Grant No. 26541059. A.-X.C. is
supported by NSFC under Grants No. 11365009. Y.-X.L. is supported by the
National Basic Research Program of China(973 Program) under Grant No.
2014CB921401, the Tsinghua University Initiative Scientific Research
Program, and the Tsinghua National Laboratory for Information Science and
Technology (TNList) Cross-discipline Foundation.

\bibliographystyle{apsrev}
\bibliography{ref}

\begin{thebibliography}{99}
\bibitem{MilburnPRA97} G. J. Milburn, J. Corney, E. M. Wright, and D. F.
Walls, Quantum dynamics of an atomic Bose-Einstein condensate in a
double-well potential, Phys. Rev. A~\textbf{55}, 4318 (1997).

\bibitem{SmerziPRL97} A. Smerzi, S. Fantoni, S. Giovanazzi, and S. R.
Shenoy, Quantum Coherent Atomic Tunneling between Two Trapped Bose-Einstein
Condensates, Phys. Rev. Lett.~\textbf{79}, 4950 (1997).

\bibitem{RaghavanPRA99} S. Raghavan, A. Smerzi, S. Fantoni, and S. R.
Shenoy, Coherent oscillations between two weakly coupled Bose-Einstein
condensates: Josephson effects, $\pi$ oscillations, and macroscopic quantum
self-trapping, Phys. Rev. A~\textbf{59}, 620 (1999).

\bibitem{MarinoPRA99} I. Marino, S. Raghavan, S. Fantoni, S. R. Shenoy, and
A. Smerzi, Bose-condensate tunneling dynamics: Momentum-shortened pendulum
with damping, Phys. Rev. A~\textbf{60}, 487 (1999).

\bibitem{GiovanazziPRL00} S. Giovanazzi, A. Smerzi, and S. Fantoni,
Josephson Effects in Dilute Bose-Einstein Condensates, Phys. Rev. Lett.~%
\textbf{84}, 4521 (2000).

\bibitem{SarchiPRA08} D. Sarchi, I. Carusotto, M. Wouters, and V. Savona,
Coherent dynamics and parametric instabilities of microcavity polaritons in
double-well systems, Phys. Rev. B~\textbf{77}, 125324 (2008).

\bibitem{BackhausNat98} S. Backhaus, S. Pereverzev, R. W. Simmonds, A.
Loshak, J. C. Davis, and R. E. Packard, Discovery of a metastable $\pi$%
-state in a superfluid $^3$He weak link, Nature (London)~\textbf{392}, 687
(1998).

\bibitem{AlbiezPRL05} M. Albiez, R. Gati, J. F\"{o}lling, S. Hunsmann, M.
Cristiani, and M. K. Oberthaler, Direct Observation of Tunneling and
Nonlinear Self-Trapping in a Single Bosonic Josephson Junction, Phys. Rev.
Lett.~\textbf{95}, 010402 (2005).

\bibitem{LevyNat07} S. Levy, E. Lahoud, I. Shomroni, and J. Steinhauer, The
a.c. and d.c. Josephson effects in a Bose-Einstein condensate, Nature
(London)~\textbf{449}, 579 (2007).

\bibitem{TrenkwalderNPY16} A. Trenkwalder, G. Spagnolli, G. Semeghini, S. Coop, M. Landini, P. Castilho, L. Pezz\`{e}, G. Modugno, M. Inguscio, A. Smerzi, and M. Fattori, Quantum phase transitions with parity-symmetry breaking and hysteresis, Nature Phys.~\textbf{12}, 826 (2016).


\bibitem{LagoudakisPRL10} K. G. Lagoudakis, B. Pietka, M. Wouters, R. Andr%
\'{e}, and B. Deveaud-Pl\'{e}dran, Coherent Oscillations in an
Exciton-Polariton Josephson Junction, Phys. Rev. Lett.~\textbf{105}, 120403
(2010).

\bibitem{AbbarchiNPy13} M. Abbarchi, A. Amo, V. G. Sala, D. D. Solnyshkov,
H. Flayac, L. Ferrier, I. Sagnes, E. Galopin, A. Lema\^{\i}tre, G. Malpuech,
and J. Bloch, Macroscopic quantum self-trapping and Josephson oscillations
of exciton polaritons, Nature Phys.~\textbf{9}, 275 (2013).

\bibitem{ACJiPRL09} A. C. Ji, Q. Sun, X. C. Xie, and W. M. Liu, Josephson
Effect for Photons in Two Weakly Linked Microcavities, Phys. Rev. Lett.~%
\textbf{102}, 023602 (2009).

\bibitem{DidierPRB11} N. Didier, S. Pugnetti, Y. M. Blanter, and R. Fazio,
Detecting phonon blockade with photons, Phys. Rev. B~\textbf{84}, 054503
(2011).

\bibitem{VoronovaPRL15} N. S. Voronova, A. A. Elistratov, and Yu. E.
Lozovik, Detuning-Controlled Internal Oscillations in an Exciton-Polariton
Condensate, Phys. Rev. Lett.~\textbf{115}, 186402 (2015).

\bibitem{RahmaniSR16} A. Rahmani and F. P. Laussy, Polaritonic Rabi and
Josephson Oscillations. Sci. Rep.~\textbf{6}, 28930 (2016).

\bibitem{LarsonPRA11} J. Larson and M. Horsdal, Photonic Josephson effect,
phase transitions, and chaos in optomechanical systems, Phys. Rev. A~\textbf{%
84}, 021804(R) (2011).

\bibitem{JHTengJPB12} J. H. Teng, S. L. Wu, B. Cui, and X X Yi, Quantum
optomechanics with quadratic cavity--membrane couplings, J. Phys. B: At.
Mol. Opt. Phys.~\textbf{45}, 185506 (2012).

\bibitem{BarzanjehPRA16} Sh. Barzanjeh and D. Vitali, Phonon Josephson
junction with nanomechanical resonators, Phys. Rev. A~\textbf{93}, 033846
(2016).

\bibitem{GeraceNPy09} D. Gerace, H. E. T\"{u}reci, A. Imamoglu, V.
Giovannetti, and R. Fazio, The quantum-optical Josephson interferometer,
Nature Phys.~\textbf{5}, 281 (2009).

\bibitem{HartmannLPR08} M. J. Hartmann, F. G. S. L. Brand\~{a}o, and M. B.
Plenio, Quantum many-body phenomena in coupled cavity arrays, Laser Photon.
Rev.~\textbf{2}, 527 (2008).


\bibitem{FollingNat07} S. F\"{o}lling, S. Trotzky, P. Cheinet, M. Feld, R. Saers, A.Widera, T. M\"{u}ller, and I. Bloch, Direct observation of second-order atom tunnelling, Nature (London)~\textbf{448}, 1029 (2007).
\bibitem{ZollnerPRL08} S. Z\"{o}llner, H.-D. Meyer, and P. Schmelcher, Few-Boson Dynamics in Double Wells: From Single-Atom to Correlated Pair Tunneling, Phys. Rev. Lett.~\textbf{100}, 040401 (2008).
\bibitem{JQLiangPRA09} J.-Q. Liang, J.-L. Liu, W.-D. Li, and Z.-J. Li, Atom-pair tunneling and quantum phase transition in the strong-interaction regime, Phys. Rev. A~\textbf{79}, 033617 (2009).
\bibitem{RubeniPRA17} D. Rubeni, J. Links, P. S. Isaac, and A. Foerster, Two-site Bose-Hubbard model with nonlinear tunneling: Classical and quantum analysis, Phys. Rev. A~\textbf{95}, 043607 (2017).

\bibitem{PietraszewiczPRA12} J. Pietraszewicz, T. Sowi\'{n}ski, M. Brewczyk,
J. Zakrzewski, M. Lewenstein, and M. Gajda, Two-component Bose-Hubbard model
with higher-angular-momentum states, Phys. Rev. A~\textbf{85}, 053638 (2012).


\bibitem{AlexanianPRA10} M. Alexanian, Scattering of two coherent photons
inside a one-dimensional coupled-resonator waveguide, Phys. Rev. A~\textbf{81%
}, 015805 (2010).

\bibitem{AlexanianPRA11} M. Alexanian, Two-photon exchange between two
three-level atoms in separate cavities, Phys. Rev. A~\textbf{83}, 023814
(2011).

\bibitem{YLDongPRA12} Y. L. Dong, S. Q. Zhu, and W. L. You, Quantum-state
transmission in a cavity array via two-photon exchange, Phys. Rev. A~\textbf{%
85}, 023833 (2012).

\bibitem{HardalJOSAB12} A. \"{U}. C. Hardal and \"{O}. E. M\"{u}stecapl{\i}o%
\u{g}lu, Transfer of spin squeezing and particle entanglement between atoms
and photons in coupled cavities via two-photon exchange, J. Opt. Soc. Am. B~%
\textbf{29}, 1822 (2012).

\bibitem{HardalJOSAB14} A. \"{U}. C. Hardal and \"{O}. E. M\"{u}stecapl{\i}o%
\"{O}lu, Spin squeezing, entanglement, and coherence in two driven,
dissipative, nonlinear cavities coupled with single- and two-photon
exchange, J. Opt. Soc. Am. B~\textbf{31}, 1402 (2014).

\bibitem{TaianarX17} G. Taian, A. V. Dodonov, Two-photon exchange
interaction from Tavis-Cummings Hamiltonian under parametric modulation,
arXiv:1703.00836 [quant-ph].

\bibitem{HWangarX17} H. Wang, S. Masis, R. Levi, O. Shtempluk, and E. Buks,
Off resonance coupling between a cavity mode and an ensemble of driven
spins, arXiv:1703.03311 [quant-ph].


\bibitem{AspelmeyerARX13} M. Aspelmeyer, T. J. Kippenberg, and F. Marquardt, Cavity Optomechanics, Rev. Mod. Phys.~\textbf{86}, 1391 (2014).


\bibitem{LudwigPRA10} M. Ludwig, K. Hammerer, and F. Marquardt, Entanglement of mechanical oscillators coupled to a nonequilibrium environment, Phys. Rev. A~\textbf{82}, 012333 (2010).
\bibitem{SeokPRA12} H. Seok, L. F. Buchmann, S. Singh, and P. Meystre, Optically mediated nonlinear quantum optomechanics, Phys. Rev. A~\textbf{86}, 063829 (2012).
\bibitem{SeokPRA13} H. Seok, L. F. Buchmann, E. M. Wright, and P. Meystre, Multimode strong-coupling quantum optomechanics, Phys. Rev. A~\textbf{88}, 063850 (2013).
\bibitem{BuchmannPRA15} L. F. Buchmann and D. M. Stamper-Kurn, Nondegenerate multimode optomechanics, Phys. Rev. A~\textbf{92}, 013851 (2015).
\bibitem{XWXuPRA13} X. W. Xu, Y. J. Zhao, and Y. X. Liu, Entangled-state engineering of vibrational modes in a multimembrane optomechanical system, Phys. Rev. A~\textbf{88}, 022325 (2013).

\bibitem{ThompsonNat08} J. D. Thompson, B. M. Zwickl, A. M. Jayich, F.
Marquardt, S. M. Girvin, and J. G. E. Harris, Strong dispersive coupling of
a high-finesse cavity to a micromechanical membrane, Nature (London)~\textbf{%
452}, 72 (2008).



\bibitem{EichenfieldNat09} M. Eichenfield, R. Camacho, J. Chan, K. J.
Vahala, and O. Painter, Optomechanical crystals, Nature (London)~\textbf{459}%
, 550 (2009).

\bibitem{MeenehanPRX15} S. M. Meenehan, J. D. Cohen, G. S. MacCabe, F. Marsili, M. D. Shaw, and O. Painter, Pulsed Excitation Dynamics of an Optomechanical Crystal Resonator near Its Quantum Ground State of Motion, Phys. Rev. X~\textbf{5}, 041002 (2015).

\bibitem{ParaisoPRX15} T. K. Para\"{\i}so, M. Kalaee, L. Zang, H. Pfeifer, F. Marquardt, and O. Painter, Position-Squared Coupling in a Tunable Photonic Crystal Optomechanical Cavity, Phys. Rev. X~\textbf{5}, 041024 (2015).

\bibitem{JahnePRA09} K. J\"{a}hne, C. Genes, K. Hammerer, M. Wallquist, E. S. Polzik, and P. Zoller, Cavity-assisted squeezing of a mechanical oscillator, Phys. Rev. A~\textbf{79}, 063819 (2009).

\bibitem{XWXuPRA15} X. W. Xu, Y. X. Liu, C. P. Sun, and Y. Li, Mechanical PT symmetry in coupled optomechanical systems, Phys. Rev. A~\textbf{92}, 013852 (2015).


\bibitem{OkamotoNP13}  H. Okamoto, A. Gourgout, C. Y. Chang, K. Onomitsu, I. Mahboob, E. Y. Chang, and H. Yamaguchi, Coherent phonon manipulation in coupled mechanical resonators, Nature Phys.~\textbf{9}, 480 (2013).
\bibitem{PHuangPRL13}  P. Huang, P. F. Wang, J. W. Zhou, Z. X. Wang, C. Y. Ju, Z. M. Wang, Y. Shen, C. K. Duan, and J. F. Du, Demonstration of Motion Transduction Based on Parametrically Coupled Mechanical Resonators, Phys. Rev. Lett.~\textbf{110}, 227202 (2013).

\bibitem{VitaliPRL07} D. Vitali, S. Gigan, A. Ferreira, H. R. B\"{o}hm, P. Tombesi, A. Guerreiro, V. Vedral, A. Zeilinger, and M. Aspelmeyer, Optomechanical Entanglement between a Movable Mirror and a Cavity Field, Phys. Rev. Lett.~\textbf{98}, 030405 (2007).
\bibitem{PalomakiNature13} T. A. Palomaki, J. W. Harlow, J. D. Teufel, R. W. Simmonds, and K. W. Lehnert, Coherent state transfer between itinerant microwave fields and a mechanical oscillator, Nature (London)~\textbf{495}, 210 (2013).
\bibitem{CohenNature15} J. D. Cohen, S. M. Meenehan, G. S.MacCabe, S. Groblacher, A. H. Safavi-Naeini, F.Marsili, M. D. Shaw, and O. Painter, Phonon counting and intensity interferometry of a nanomechanical resonator, Nature (London)~\textbf{520}, 522 (2015).

\end{thebibliography}

\end{document}